\documentclass[reprint,aps,prl,twocolumn,superscriptaddress,nofootinbib,amsmath,amssymb]{revtex4-2}
\usepackage{graphicx,xcolor}
\usepackage[colorlinks=true,linkcolor=blue!50!black,citecolor=blue!50!black,urlcolor=blue!50!black]{hyperref}
\usepackage{aas_macros}
\usepackage{amsmath}
\usepackage{amssymb}
\usepackage{amsfonts}
\usepackage{physics}
\usepackage[normalem]{ulem}
\usepackage[english]{babel}
\usepackage[utf8]{inputenc}

\newcommand{\Ain}{A_{\rm in}}
\newcommand{\upk}{u_{\rm peak}}
\newcommand{\sch}{Schwarzschild }

\begin{document}

\title{The Start Time of Ringdown}

\author{Adrien Kuntz}
\affiliation{CENTRA, Departamento de Física, Instituto Superior Técnico – IST,
Universidade de Lisboa – UL, Avenida Rovisco Pais 1, 1049-001 Lisboa, Portugal}

\begin{abstract}
We determine the convergence of quasi-normal modes sums for a point-particle plunging into a Schwarzschild black hole. While a dynamical amplitude model reproduces the waveform after the particle crosses the bounce radius, a constant amplitude sum converges only after the start time $t_S$, which corresponds to the advanced time at which the particle hits the central singularity. For a plunge from the ISCO, $t_S \simeq 9.4 M$ after the peak of the $22$ mode.  After $t_S$, dynamical and constant amplitudes models are equivalent; our analysis also implies that any ``direct wave'' must reduce to the late-time tail after $t_S$.
\end{abstract}

\maketitle

\textit{\textbf{Introduction.}} The last stage of gravitational waves (GW) of a binary black hole (BH) coalescence, the ringdown, is traditionally modeled with the proper modes of oscillation of a BH, the quasi-normal modes (QNMs)~\cite{Berti:2025hly}. Measuring several of them tests General Relativity (GR), and the loud event GW250114 allowed the measurement of two modes, one of them being an overtone~\cite{LIGOScientific:2025rid,LIGOScientific:2025wao}. 
An outstanding open issue in this field concerns the regime of validity of ringdown: from what time on can we model the waveform with a sum of damped sinusoids of constant amplitude~\cite{Nollert:1998ys,Szpak:2004sf,Chavda:2024awq,Oshita:2024wgt,Bourg:2026clj}? This ``start time'' uncertainty enters all the ringdown analyses of the LIGO/Virgo/KAGRA (LVK) collaboration~\cite{LIGOScientific:2020tif, LIGOScientific:2026wpt, LIGOScientific:2021sio, LIGOScientific:2019fpa, LIGOScientific:2025wao} and fueled debates in the community~\cite{Giesler:2019uxc,Bhagwat:2019dtm,Baibhav:2023clw,Cotesta:2022pci}.  In this Letter, we predict the value of the start time for a point-particle plunging into a non-rotating BH, and relate it to properties of the particle's trajectory. We furthermore give the rate of convergence of QNM sums to the waveform as a function of the number of overtones included in the model, and prove that the signal is exactly given by a sum of QNMs of constant amplitudes together with the late-time tail after the start time.

\textbf{\textit{Setup.}} We consider a non-rotating BH in \sch coordinates $(t,r,\theta,\phi)$, and use units $G=c=2M=1$, with $M$ the black-hole mass. We define the tortoise coordinate $x=r+\log(r-1)$, retarded and advanced times $u=t-x$ and $v=t+x$. A particle of mass $\mu\ll M$ follows a geodesic plunging into the BH, whose trajectory $x_p(t)$ has energy $E$ and angular momentum $L$; we normalize time so that the particle crosses the light-ring $r=3M$ at $t_\mathrm{LR}=0$ ($u_\mathrm{LR}\simeq -1.61M$). Special attention will be devoted to a particular trajectory~\cite{Ori:2000zn, Hadar:2009ip}, where the particle plunges from the Innermost Stable Circular Orbit (ISCO) at $r=6M$, with $E=2\sqrt2/3$ and $L=2\sqrt3M$; an explicit expression of $x_p(t)$ in this case is given in the Supplemental Material (SM). 
We neglect radiation reaction. We decompose the GW strain into multipoles $h_{\ell m}$, where $r h_{\ell m}$ is proportional to a master function $\Psi$, of Zerilli type for even $\ell+m$ and of Regge--Wheeler type for odd $\ell+m$, which obeys
\begin{align}
&(\partial_t^2-\partial_x^2+V)\Psi= S(t,x) \; ,\label{eq:rwz} \\
&S(t,x) = F(t)\,\delta(x-x_p(t))+G(t)\, \partial_x\delta(x-x_p(t)) \; , \label{eq:source}
\end{align}
where $V$ is the corresponding potential, and $F$, $G$ are known functions of time along
the trajectory~\cite{Martel:2005ir, Nagar:2005ea}. In the following it will be useful to change variables and write
\begin{equation} \label{eq:sourceu}
    S(u,x) = F^u(u) \,\delta(x-x_p(u))+G^u(u)\, \partial_x\delta(x-x_p(u)) ,
\end{equation}
where now the $x$ derivative is taken at fixed $u$. 
Using properties of the Dirac delta function, we get $G^u = G/(1-\dot x_p)^2$, $F^u = F/(1-\dot x_p)-\dot G/(1-\dot x_p)^2-G \ddot x_p/(1-\dot x_p)^3 $ where a dot denotes time derivative; we similarly define $S(v,x)$. We solve Eq.~\eqref{eq:rwz} in the time domain on hyperboloidal slices~\cite{Zenginoglu:2010cq,Bernuzzi:2011aj}, which gives the waveform at null infinity with a relative accuracy $\sim 10^{-5}$; additional details are given in the SM. We focus on $\ell m=22$, and measure time from the peak of $|h_{22}|$, at retarded time $\upk$.

\textit{\textbf{Dynamical QNM amplitudes.}} For an observer at null infinity, the retarded Green function of Eq.~\eqref{eq:rwz} is~\cite{Leaver:1986gd}
\begin{equation}
\mathcal G(u;t',x')=i\int_{-\infty + i \gamma}^{\infty + i \gamma}\frac{\mathrm{d}\omega}{2\pi}\,e^{-i\omega(u-t')}\,\frac{u_{\rm in}(x',\omega)}
{2\omega\Ain(\omega)} \; ,\label{eq:green}
\end{equation}
where $\gamma>0$, $u_{\rm in}$ is the solution of the homogeneous equation that is ingoing at the horizon, $u_{\rm in}\simeq
e^{-i\omega x}$ for $x\to-\infty$, and $u_{\rm in}\simeq\Ain e^{-i\omega x}+A_{\rm out}e^{i\omega x}$ for $x\to\infty$. The QNM frequencies $\omega_n$ are the zeros of $\Ain$ (overtones $n=0,1,\dots$, and their mirrors $-\bar\omega_n$, all located in the lower half complex plane). With this definition there is no overtone n = 8: it would sit at the algebraically special frequency $\omega=-4i$, where $\Ain$ has neither a zero nor a pole~\cite{MaassenvandenBrink:2000iwh}.

\begin{figure*}
\centering\includegraphics[width=\textwidth]{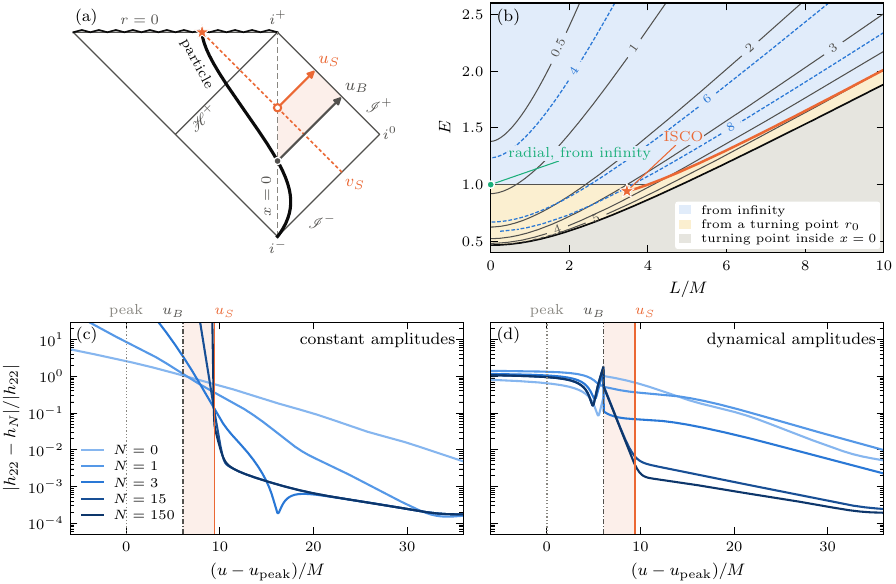}
\caption{QNM start time for a particle plunging into a Schwarzschild BH. (a)~Penrose diagram of a \sch BH, with the worldline of a particle plunging into the
singularity $r=0$. The particle crosses the bounce radius $x_B=0$ (dashed), from which an outgoing light ray
reaches the observer at the bounce time $u_B$, and then hits the singularity at the advanced time $v_S$ (star). The ingoing
ray $v=v_S$ (orange dashed) crosses $x_B=0$ at the mirror image of the star under $U\to-U$ (open circle), from which an
outgoing ray reaches the observer at the start time $u_S=v_S$. (b)~start time for plunging geodesics of energy $E$
and angular momentum $L$: contours of $(u_S-u_B)/M$ (solid) and, for the plunges that cross the light ring, of
$(u_S-u_{\rm LR})/M$ (dashed). The particle comes from infinity in the blue region ($R(r)>0$ for all $r$), and otherwise
from the turning point $r_0$ closest to the horizon ($R(r_0)=0$); in the grey region $r_0<r_B$, and the particle does
not cross the bounce radius. Orange line: critical plunges, which start from an unstable circular orbit; star:
ISCO plunge; dot: radial plunge from rest at infinity. (c),~(d)~Relative error of the QNM sums truncated at $N$
overtones (and their mirrors) for the $\ell m=22$ waveform of the ISCO plunge, with constant amplitudes $\tilde C_n$
(including the algebraically special mode $n=8$ for $N\geq8$, see the SM) and with dynamical amplitudes $C_n(u)$.
Vertical lines: peak of $|h_{22}|$ (dotted), $u_B$ (dash-dotted) and $u_S$ (orange). The shading marks $u_B<u<u_S$
in (a), (c) and (d).}
\label{fig:conv}
\end{figure*}

Recent works building on the seminal article from Leaver~\cite{Leaver:1986gd} obtained an explicit decomposition of the \sch Green function~\eqref{eq:green} into QNMs, late-time tail and prompt response~\cite{Kuntz:2025gdq, Arnaudo:2025uos,Arnaudo:2025kit,Ma:2026qbq,DeAmicis:2025xuh,DeAmicis:2026wqd, DeAmicis:2026tus,Su:2026fvj}. In this article we focus on the QNM part, which is activated for times satisfying the \textit{bounce condition} $u \geq t' + |x'|$~\cite{Arnaudo:2025uos,DeAmicis:2026wqd,Arnaudo:2026tcy} and defines the convergence region of the QNM sum of overtones in the Green function~\eqref{eq:green}:
\begin{equation}
\mathcal G_{\rm QNM}= \Theta(u-t' - |x'|) \sum_n \, \frac{e^{-i\omega_n(u-t')}u_n(x')}{2\omega_n\Ain'(\omega_n)} \; , \label{eq:gqnm}
\end{equation}
with $u_n=u_{\rm in}(x,\omega_n)$. The sum in Eq.~\eqref{eq:gqnm} includes all overtones as well as mirror modes, whose frequency has a negative real part, $\omega_n^\mathrm{mirror} = - (\omega_n)^*$. Physically, the bounce condition can be seen as the time it takes for a signal emitted at $(t', x')$ to travel to the bounce radius $x_B=0$ ($r_B\simeq2.56M$) and then out to the observer~\cite{Arnaudo:2026tcy}. We numerically obtain the QNM frequencies $\omega_n$ and wavefunctions $u_n$ up to $n=150$ using the Leaver algorithm~\cite{Leaver:1985ax}; on the other hand, we use the method of Ref.~\cite{Lo:2025njp} to obtain $A_\mathrm{in}(\omega)$ and its derivatives at the QNM frequencies, see the SM for details. 
By convolving the Green function~\eqref{eq:gqnm} with the source~\eqref{eq:source}, we can obtain the QNM part of the time-domain signal which defines dynamical QNM amplitudes $C_n(u)$~\cite{DeAmicis:2025xuh,DeAmicis:2026wqd}:
\begin{align}
 \Psi_{\rm QNM}(u)&=\sum_nC_n(u)\,e^{-i\omega_nu} \, ,  \label{eq:psiQNM}\\
C_n(u)&=\int_{t'+|x'|<u}\mathrm{d}t'\mathrm{d}x'\, \frac{e^{i\omega_nt'}u_n(x')}{2\omega_n\Ain'(\omega_n)} \,S(t',x') \; ,\label{eq:Cn_def}
\end{align}
The particle crosses $x=0$, where $u=v=t$, at the bounce time $u_B$; for the ISCO plunge, $u_B - u_\mathrm{LR}\simeq 5.24M$. For $u<u_B$, every point on the source trajectory in the past light-cone of the observer satisfies $x_p(t')>0$, so that $t'+|x_p(t')| = t'+x_p(t')$. On the other hand, for $u>u_B$, every point on the past light-cone also satisfies $t'+|x_p(t')|<u$: hence in this case the domain of integration of Eq.~\eqref{eq:Cn_def} is set by causality alone, $t'-x_p(t')<u$ (see Fig.~\ref{fig:conv}(a)). 
This feature suggests to use the change of variable $t' \to v'$ in the integral~\eqref{eq:Cn_def} for $u<u_B$, and $t' \to u'$ for $u>u_B$. This trick avoids the presence of the so-called impulsive terms in the waveform~\cite{DeAmicis:2025xuh, DeAmicis:2026wqd, Kuntz:2026xep}, because the delta-function derivative in Eq.~\eqref{eq:source} does not act on the boundary of the integral. We get, for $u>u_B$:
\begin{align} \label{eq:Cnu>ub}
    &C_n(u) = \int_{-\infty}^u \mathrm{d}u' \, \frac{e^{i \omega_n u'}}{2 \omega_n A'_\mathrm{in}(\omega_n)} \bigg(  F^u  \lambda_n^u - G^u \partial_x \lambda_n^u\bigg) \; ,
\end{align}
where the term in parenthesis is evaluated at the particle location $x'=x_p(u')$, $\lambda_n^u(x) = e^{i \omega_n x} u_n(x)$, and $F^u$ and $G^u$ have been defined in Eq.~\eqref{eq:sourceu}. 
Similarly, for $u<u_B$,
\begin{align} \label{eq:Cnu<ub}
    &C_n(u) = \int_{-\infty}^u \mathrm{d}v' \, \frac{e^{i \omega_n v'}}{2 \omega_n A'_\mathrm{in}(\omega_n)} \bigg(  F^v  \lambda_n^v - G^v \partial_x \lambda_n^v \bigg) \; ,
\end{align}
with $\lambda_n^v(x) = e^{-i \omega_n x} u_n(x)$. 
At late times $u \rightarrow \infty$, $C_n(u)$ may diverge because the decay of the source is not enough to compensate the growth of the $e^{i \omega_n u'}$ term in~\eqref{eq:Cnu>ub}; one can check that this happens for all $n \geq 1$~\cite{Leaver:1986gd}. A regularization procedure is thus needed to extract a physical constant QNM amplitude at late times.

\textit{\textbf{Regularization.}} Near the horizon, $u'\to\infty$, the source along the orbit is a regular function of the Kruskal coordinate $U=-e^{-u/2}$, since the worldline crosses the horizon $U=0$ smoothly. We expand the integrand of Eq.~\eqref{eq:Cnu>ub} in the variable $U$; this series which converges for $u>u_S$, the ringdown start time, that we leave undetermined for now. The physical picture developed in~\cite{DeAmicis:2026wqd} suggests that $u_S>u_B$, a fact that we will recover in the following. Choosing an arbitrary intermediate time $u_1>u_S$, we can write $C_n(u)$ in Eq.~\eqref{eq:Cnu>ub} as~\cite{Kuntz:2026xep}
\begin{equation}
C_n(u)= C_n(u_1)+ \int_{u_1}^u \mathrm{d}u' \; e^{i \omega_n u'} \sum_{k\geq1} \frac{a_k(\omega_n)\,e^{-k u'/2}}{2 \omega_n A'_\mathrm{in}(\omega_n)} ,\label{eq:ak}
\end{equation}
where we obtain coefficients $a_k(\omega_n)$ from the near-horizon expansion of the integrand. The integral becomes solvable in closed-form; the vanishing of the so-called redshift modes~\cite{Mino:2008at,Zimmerman:2011dx,DeAmicis:2025xuh, DeAmicis:2026wqd,Kuntz:2026xep} corresponds to the statement that the upper bound of the integral vanishes once summing all overtones in the waveform, up to a very small residual related to the branch-cut contribution that we will neglect. This statement holds true apart from the algebraically special frequency $k=8$ ($\omega=-4i$): see Ref.~\cite{Kuntz:2026xep} and the SM, in which we deal with the $k=8$ special case, for more details. The vanishing of redshift modes allows us to define the regularized \textit{constant} QNM amplitude $\tilde C_n$, obtained by throwing away the upper bound of the integral in Eq.~\eqref{eq:ak}:
\begin{equation}
    \tilde C_n = C_n(u_1) - \sum_{k \geq 1} \frac{a_k(\omega_n) e^{(i \omega_n - k/2)u_1}}{2 \omega_n A'_\mathrm{in}(\omega_n)(i \omega_n - k/2)} \; .
\end{equation}
The regularization procedure used by Detweiler and Szedenits and in many subsequent works~\cite{Detweiler:1979xr,Leaver:1986gd,Hadar:2009ip, Berti:2006wq,Zhang:2013ksa,DellaRocca:2025zbe,Kuchler:2025hwx} just corresponds to send $u_1 \rightarrow \infty$: in this case the terms with $k \leq 2 |\mathrm{Im} \omega_n|$ regulate the divergences of the integral in $C_n(u_1)$, while higher $k$'s converge to zero. Our derivation makes it clear that any other value of $u_1$ gives the same value of $\tilde C_n$, as long as the series expansion in Eq.~\eqref{eq:ak} converges at the intermediate time $u_1$, i.e. $u_1>u_S$. We have numerically computed $\tilde C_n$ up to $n=150$ for a plunge from the ISCO and checked it reproduces published results, available only for low overtones~\cite{Hadar:2009ip,Folacci:2018cic}. Our values for $\tilde C_n$ are available online~\cite{repo}.

\textit{\textbf{Convergence and start time.}} When summing a large number of overtones, the dynamical QNM model converges to the full waveform only for $u>u_B$ (see Fig.~\ref{fig:conv}(d)): this is to be expected since at earlier times the prompt response also contributes to the waveform~\cite{Ma:2026qbq, DeAmicis:2026tus}. In order to determine the rate of convergence, we can notice that at large overtone numbers one has $\lambda_n^u \rightarrow 1$ for $x<0$, $\omega_n \propto n$ and $\Ain'(\omega_n) = \mathcal{O}(1)$~\cite{Kuntz:2026xep} so that the integral defining $C_n$ for $u > u_B$ is given by $C_n(u) \propto {\omega_n}^{-1} \int_{-\infty}^u e^{i \omega_n u'} F^u(u') \mathrm{d}u' $. A simple integration by parts then gives $C_n(u) e^{-i \omega_n u} \propto n^{-2}$, and the remainder in Eq.~\eqref{eq:psiQNM} after truncating the model to a number $N$ of overtones scales as $1/N$. 
The remaining small discrepancy between the dynamical QNM model and the numerical waveform for $u>u_B$ can be attributed to the late-time tail, which originates from the branch-cut contribution to Eq.~\eqref{eq:green} and from an algebraically special mode which is not a zero of $A_\mathrm{in}$ but a pole of $u_\mathrm{in}$ in the even sector, see the SM for more details. 

The constant-amplitude sum behaves differently. Its convergence is set by the singularity of the source nearest to $U=0$ of the source along the worldline, continued into the BH: this corresponds to $r=0$. Introducing the advanced Kruskal coordinate $V=e^{v/2}$ and noting that $U_S V_S=1$ at the singularity, we see that $U_S = e^{-v_S/2}$ in the interior region of spacetime, where $v_S$ is the value of the advanced time coordinate at which the particle hits the singularity. This translates into a convergence time $u_S = v_S$ in the physical exterior region, see Fig.~\ref{fig:conv}(a). Since $v$ increases along any plunging worldline and equals $u_B$ at $x=0$, $u_S>u_B$ for any plunge: constant QNMs necessarily converge after the bounce time. Since $\mathrm{d}x_p/\mathrm{d}t = -\sqrt{R}/E $ for a \sch geodesic, with $R(r) = E^2-(1-1/r)(1+L^2/r^2)$, we get the simple result
\begin{equation}
    u_S-u_B = v_S - v_B = \int_0^{r_B} \mathrm{d}r \, \frac{1 + L^2/r^2}{\sqrt{R(r)}(E+\sqrt{R(r)})} \; , \label{eq:startTime}
\end{equation}
which gives the dependency of the start time on the orbit parameters. For the ISCO plunge, $u_S \simeq u_B + 3.37M \simeq u_{\rm LR} + 8.61M \simeq \upk + 9.4M$,  comparable with the results of Ref.~\cite{Bourg:2026clj} and to the times from which constant-amplitude fits are usually found to be reliable~\cite{Bhagwat:2019dtm,Baibhav:2023clw,Cheung:2023vki}. For a radial infall from infinity, $u_S \simeq u_B + 0.87M \simeq u_{\rm LR} + 4.38M$. In the ultra-relativistic limit $E \rightarrow \infty$ at fixed $L$, the trajectory approaches a null geodesic $\mathrm{d}x_p/\mathrm{d}t \rightarrow -1$ while the start time $u_S \rightarrow u_B$: this is consistent with the interpretation of the bounce radius as being the QNM convergence region of a null ray falling into the BH~\cite{Arnaudo:2026tcy}. We show in Fig.~\ref{fig:conv}(b) the dependence of $u_S$ on the parameters of the orbit.

The constant-amplitude sum therefore diverges before $u_S$, where adding overtones makes it worse, and converges exponentially after it, each overtone being suppressed by $e^{-|{\rm Im}\,\omega_n|(u-u_S)}$ (Fig.~\ref{fig:conv}(c)). 
The dynamical amplitudes regroup each overtone with its own redshift modes, and turn this divergent sum into a convergent one between $u_B$ and $u_S$. Since dynamical and constant amplitudes sums converge to the same limit after $u_S$, the residual between the constant-amplitude sum and the numerical waveform visible in Fig.~\ref{fig:conv}(c) is also given by the late-time tail. 
Let us emphasize that, even if the start time seems to depend on the interior geometry of the BH, it is fixed by the analytic continuation of the exterior trajectory: modifying the interior geometry only (for instance near $r=0$) would not change it. 


\textit{\textbf{Discussion.}} In this Letter we have found the time at which a QNM expansion in overtones converge for a point-particle plunging in a \sch BH. We hope our results will be useful for building more accurate ringdown templates as well as to establish more stringent tests of GR: since it is now possible to relate the start time to properties of the BH, it can be used as a way to test the nature of the remnant itself. We also discovered the interesting feature that constant QNM models converge much faster than dynamical QNMs after $u_S$, so that a constant amplitude model with few overtones should be more accurate a few $M$ after $u_S$. Summing an infinite number of overtones, both models converge to the same waveform after $u_S$. Nonetheless, we have uncovered a portion of the signal between the bounce and the start time for which the signal is a pure dynamical QNM (together with the branch cut) with no prompt response, and fitting a constant QNM model there would not recover the physical amplitudes $\tilde C_n$. 


Our results are also relevant to the ``direct wave'' models, see e.g.~\cite{Oshita:2025qmn, Weller:2026wjh}. A broad definition of a direct wave is whatever remains in the signal after removing constant amplitude QNMs. The results of this article make it clear that a direct wave should reduce to the late-time tail after $u_S$, and is therefore not a probe of the near-horizon geometry~\cite{Kankani:2026kst,Kuntz:2026xep,Dyer:2026sox}. The direct wave has always been identified in the data by using a QNM filter effectively removing constant QNMs from the signal~\cite{Ma:2022wpv}; however, the filter can also introduce other additional effects which obscure the interpretation of the direct wave signal~\cite{Cheung:2026gfd}. We think the picture developed in this Letter is clear enough to bypass the use of any filter regarding the physical interpretation of the signal. Our results hold for non-rotating BHs; the rotating case is technically more involved, but the physics involved is similar and our argument based on the singularities of the source should extend to this case as well. It would also be interesting to understand if our results could be extrapolated to comparable-mass mergers.


\begin{acknowledgments}
\textit{\textbf{Acknowledgments}} I would like to thank Marina de Amicis, Adam Pound and Ling Sun for discussions. I thank the Fundação para a Ciência e Tecnologia (FCT), Portugal, for the financial support to the Center for Astrophysics and Gravitation (CENTRA/IST/ULisboa) through grant No. UID/PRR/00099/2025 and grant No. UID/00099/2025, as well as to the FCT project ``Gravitational waves as a new probe of fundamental physics and astrophysics'' grant agreement 2023.07357.CEECIND/CP2830/CT0003. During the preparation of this work the author used Claude (Anthropic, Claude Opus 5.5, via Claude Code) to carry out and check symbolic and numerical calculations, to search the literature, and to draft and edit the text under the author’s direction. All derivations, numerical results and statements were reviewed and verified by the author, who takes full responsibility for the content of the manuscript.
\end{acknowledgments}

\bibliography{refs}

\def\SMinMain{}
\ifx\SMinMain\undefined
\documentclass[aps,prd,twocolumn,superscriptaddress,nofootinbib,amsmath,amssymb]{revtex4-2}
\usepackage{graphicx,xcolor}
\usepackage{xr-hyper}
\usepackage[colorlinks=true,linkcolor=blue!50!black,citecolor=blue!50!black,urlcolor=blue!50!black]{hyperref}
\externaldocument[M-]{main}[main.pdf]
\makeatletter       
\DeclareRobustCommand*{\SM@xref}[1]{\@ifundefined{r@#1}{\SM@ref{M-#1}}{\SM@ref{#1}}}
\AddToHook{begindocument/end}{\NewCommandCopy\SM@ref\ref\let\ref\SM@xref}
\makeatother
\renewcommand{\bibnumfmt}[1]{[S#1]}
\renewcommand{\citenumfont}[1]{S#1}
\newcommand{\SMend}{\bibliography{refs}\end{document}}

@article{Bourg:2026clj,
    author = {Bourg, Patrick and K{\"u}chler, Lorenzo and Leather, Benjamin and Cunningham, Kevin and Lewis, Jack and Pound, Adam and Bonga, B{\'e}atrice and Casals, Marc and Panosso Macedo, Rodrigo},
    title = "{Dissection of a merger-ringdown waveform in the small-mass-ratio limit}",
    eprint = "2609.06861",
    archivePrefix = "arXiv",
    primaryClass = "gr-qc",
    month = "9",
    year = "2026",
    journal=""
}

@article{Dyer:2026sox,
    author = "Dyer, Richard and Kankani, Anuj and Marks, Gareth Arturo and McWilliams, Sean T. and Moore, Christopher J.",
    title = "{Moving Beyond Quasinormal Modes and Reinterpreting Direct Waves in Binary Black Hole Merger-Ringdowns}",
    eprint = "2610.25176",
    archivePrefix = "arXiv",
    primaryClass = "gr-qc",
    month = "10",
    year = "2026",
    journal=""
}

@article{Cheung:2026gfd,
    author = "Cheung, Mark Ho-Yeuk",
    title = "{How much of the black hole ringdown is sourced within the light ring?}",
    eprint = "2608.29466",
    archivePrefix = "arXiv",
    primaryClass = "gr-qc",
    month = "8",
    year = "2026",
    journal=""
}

@article{LIGOScientific:2021sio,
    author = "Abbott, R. and others",
    collaboration = "LIGO Scientific, VIRGO, KAGRA",
    title = "{Tests of General Relativity with GWTC-3}",
    eprint = "2112.06861",
    archivePrefix = "arXiv",
    primaryClass = "gr-qc",
    reportNumber = "LIGO-P2100275",
    doi = "10.1103/PhysRevD.112.084080",
    journal = "Phys. Rev. D",
    volume = "112",
    number = "8",
    pages = "084080",
    year = "2025"
}

@article{LIGOScientific:2026wpt,
    author = "Abac, A. G. and others",
    collaboration = "LIGO Scientific, VIRGO, KAGRA",
    title = "{GWTC-4.0: Tests of General Relativity. III. Tests of the Remnants}",
    eprint = "2603.19021",
    archivePrefix = "arXiv",
    primaryClass = "gr-qc",
    reportNumber = "LIGO-P2500067",
    month = "3",
    year = "2026",
    journal=""
}

@article{Kuchler:2025hwx,
    author = {K{\"u}chler, Lorenzo and Comp{\`e}re, Geoffrey and Pound, Adam},
    title = "{Self-force framework for merger-ringdown waveforms}",
    eprint = "2506.02189",
    archivePrefix = "arXiv",
    primaryClass = "gr-qc",
    doi = "10.1088/1361-6382/ae2b44",
    journal = "Class. Quant. Grav.",
    volume = "43",
    number = "1",
    pages = "015018",
    year = "2026"
}

@article{DeAmicis:2026tus,
    author = "De Amicis, Marina and Cannizzaro, Enrico",
    title = "{Post-Minkowskian expansion of the Prompt Response in a Schwarzschild background}",
    eprint = "2601.11706",
    archivePrefix = "arXiv",
    primaryClass = "gr-qc",
    month = "1",
    year = "2026",
    journal=""
}

@article{DeAmicis:2025xuh,
    author = "De Amicis, Marina and Cannizzaro, Enrico and Carullo, Gregorio and Sberna, Laura",
    title = "{Dynamical quasinormal mode excitation}",
    eprint = "2506.21668",
    archivePrefix = "arXiv",
    primaryClass = "gr-qc",
    doi = "10.1103/bgjv-lvyd",
    journal = "Phys. Rev. D",
    volume = "113",
    number = "2",
    pages = "024048",
    year = "2026"
}

@article{DeAmicis:2026wqd,
    author = "De Amicis, Marina and Cannizzaro, Enrico and Carullo, Gregorio and Kuntz, Adrien and Sberna, Laura",
    title = "{Dynamical quasinormal mode excitation II: propagation and convergence in Schwarzschild}",
    eprint = "2605.16492",
    archivePrefix = "arXiv",
    primaryClass = "gr-qc",
    month = "5",
    year = "2026",
    journal=""
}

@article{Kuntz:2026xep,
    author = "Kuntz, Adrien and Della Rocca, Matteo",
    title = "{Vanishing of all redshift modes in Schwarzschild ringdown}",
    eprint = "2606.07173",
    archivePrefix = "arXiv",
    primaryClass = "gr-qc",
    month = "6",
    year = "2026",
    journal=""
}

@article{Kuntz:2025gdq,
    author = "Kuntz, Adrien",
    title = {{Green function of the P{\"o}schl-Teller potential}},
    eprint = "2510.17954",
    archivePrefix = "arXiv",
    primaryClass = "gr-qc",
    doi = "10.21468/SciPostPhys.20.4.120",
    journal = "SciPost Phys.",
    volume = "20",
    number = "4",
    pages = "120",
    year = "2026"
}

@article{Arnaudo:2026tcy,
    author = "Arnaudo, Paolo and Withers, Benjamin",
    title = "{Bouncing singularities in Schwarzschild: a geometric origin of the QNM convergence region}",
    eprint = "2605.16489",
    archivePrefix = "arXiv",
    primaryClass = "gr-qc",
    month = "5",
    year = "2026",
    journal=""
}

@article{Arnaudo:2025uos,
    author = "Arnaudo, Paolo and Carballo, Javier and Withers, Benjamin",
    title = "{Beyond quasinormal modes: A complete mode decomposition of black hole perturbations}",
    eprint = "2510.18956",
    archivePrefix = "arXiv",
    primaryClass = "gr-qc",
    doi = "10.1103/1bcl-q79x",
    journal = "Phys. Rev. D",
    volume = "114",
    number = "2",
    pages = "024025",
    year = "2026"
}

@article{Arnaudo:2025kit,
    author = "Arnaudo, Paolo and Withers, Benjamin",
    title = "{Price{\textquoteright}s law from quasinormal modes}",
    eprint = "2511.17703",
    archivePrefix = "arXiv",
    primaryClass = "gr-qc",
    doi = "10.1103/xlg9-nc6k",
    journal = "Phys. Rev. D",
    volume = "114",
    number = "4",
    pages = "044085",
    year = "2026"
}

@article{Mino:2008at,
    author = "Mino, Yasushi and Brink, Jeandrew",
    title = "{Gravitational Radiation from Plunging Orbits: Perturbative Study}",
    eprint = "0809.2814",
    archivePrefix = "arXiv",
    primaryClass = "gr-qc",
    doi = "10.1103/PhysRevD.78.124015",
    journal = "Phys. Rev. D",
    volume = "78",
    pages = "124015",
    year = "2008"
}

@article{Zimmerman:2011dx,
    author = "Zimmerman, Aaron and Chen, Yanbei",
    title = "{New Generic Ringdown Frequencies at the Birth of a Kerr Black Hole}",
    eprint = "1106.0782",
    archivePrefix = "arXiv",
    primaryClass = "gr-qc",
    doi = "10.1103/PhysRevD.84.084012",
    journal = "Phys. Rev. D",
    volume = "84",
    pages = "084012",
    year = "2011"
}

@article{Leaver:1985ax,
    author = "Leaver, E. W.",
    title = "{An Analytic representation for the quasi normal modes of Kerr black holes}",
    doi = "10.1098/rspa.1985.0119",
    journal = "Proc. Roy. Soc. Lond. A",
    volume = "402",
    pages = "285--298",
    year = "1985"
}

@article{Leaver:1986gd,
    author = "Leaver, Edward W.",
    title = "{Spectral decomposition of the perturbation response of the Schwarzschild geometry}",
    doi = "10.1103/PhysRevD.34.384",
    journal = "Phys. Rev. D",
    volume = "34",
    pages = "384--408",
    year = "1986"
}

@article{Nollert:1993zz,
    author = "Nollert, Hans-Peter",
    title = "{Quasinormal modes of Schwarzschild black holes: The determination of quasinormal frequencies with very large imaginary parts}",
    doi = "10.1103/PhysRevD.47.5253",
    journal = "Phys. Rev. D",
    volume = "47",
    pages = "5253--5258",
    year = "1993"
}

@article{Nollert:1998ys,
    author = "Nollert, Hans-Peter and Price, Richard H.",
    title = "{Quantifying excitations of quasinormal mode systems}",
    eprint = "gr-qc/9810074",
    archivePrefix = "arXiv",
    doi = "10.1063/1.532698",
    journal = "J. Math. Phys.",
    volume = "40",
    pages = "980--1010",
    year = "1999"
}

@article{Berti:2025hly,
    author = "Berti, Emanuele and others",
    editor = "Berti, Emanuele and Cardoso, Vitor and Carullo, Gregorio",
    title = "{Black hole spectroscopy: from theory to experiment}",
    eprint = "2505.23895",
    archivePrefix = "arXiv",
    primaryClass = "gr-qc",
    reportNumber = "RUP-25-10; YITP-25-64; RIKEN-iTHEMS-Report-25",
    doi = "10.1088/1361-6382/ae59e2",
    journal = "Class. Quant. Grav.",
    volume = "43",
    number = "12",
    pages = "123001",
    year = "2026"
}

@article{Berti:2006wq,
    author = "Berti, Emanuele and Cardoso, Vitor",
    title = "{Quasinormal ringing of Kerr black holes. I. The Excitation factors}",
    eprint = "gr-qc/0605118",
    archivePrefix = "arXiv",
    doi = "10.1103/PhysRevD.74.104020",
    journal = "Phys. Rev. D",
    volume = "74",
    pages = "104020",
    year = "2006"
}

@article{Lo:2025njp,
    author = "Lo, Rico K. L. and Sabani, Leart and Cardoso, Vitor",
    title = "{Quasinormal modes and excitation factors of Kerr black holes}",
    eprint = "2504.00084",
    archivePrefix = "arXiv",
    primaryClass = "gr-qc",
    doi = "10.1103/PhysRevD.111.124002",
    journal = "Phys. Rev. D",
    volume = "111",
    number = "12",
    pages = "124002",
    year = "2025"
}

@article{Hadar:2009ip,
    author = "Hadar, Shahar and Kol, Barak",
    title = "{Post-ISCO Ringdown Amplitudes in Extreme Mass Ratio Inspiral}",
    eprint = "0911.3899",
    archivePrefix = "arXiv",
    primaryClass = "gr-qc",
    doi = "10.1103/PhysRevD.84.044019",
    journal = "Phys. Rev. D",
    volume = "84",
    pages = "044019",
    year = "2011"
}

@article{Folacci:2018cic,
    author = "Folacci, Antoine and Ould El Hadj, Mohamed",
    title = "{Multipolar gravitational waveforms and ringdowns generated during the plunge from the innermost stable circular orbit into a Schwarzschild black hole}",
    eprint = "1806.01577",
    archivePrefix = "arXiv",
    primaryClass = "gr-qc",
    doi = "10.1103/PhysRevD.98.084008",
    journal = "Phys. Rev. D",
    volume = "98",
    number = "8",
    pages = "084008",
    year = "2018"
}

@article{DellaRocca:2025zbe,
    author = "Della Rocca, Matteo and Pezzella, Laura and Berti, Emanuele and Gualtieri, Leonardo and Maselli, Andrea",
    title = "{Quasinormal ringing of Kerr black holes. III. Excitation coefficients for equatorial inspirals from the innermost stable circular orbit}",
    eprint = "2512.07959",
    archivePrefix = "arXiv",
    primaryClass = "gr-qc",
    doi = "10.1103/ydp5-ktpn",
    journal = "Phys. Rev. D",
    volume = "114",
    number = "4",
    pages = "044036",
    year = "2026"
}

@article{Giesler:2019uxc,
    author = "Giesler, Matthew and Isi, Maximiliano and Scheel, Mark A. and Teukolsky, Saul",
    title = "{Black Hole Ringdown: The Importance of Overtones}",
    eprint = "1903.08284",
    archivePrefix = "arXiv",
    primaryClass = "gr-qc",
    doi = "10.1103/PhysRevX.9.041060",
    journal = "Phys. Rev. X",
    volume = "9",
    number = "4",
    pages = "041060",
    year = "2019"
}

@article{Baibhav:2023clw,
    author = "Baibhav, Vishal and Cheung, Mark Ho-Yeuk and Berti, Emanuele and Cardoso, Vitor and Carullo, Gregorio and Cotesta, Roberto and Del Pozzo, Walter and Duque, Francisco",
    title = "{Agnostic black hole spectroscopy: Quasinormal mode content of numerical relativity waveforms and limits of validity of linear perturbation theory}",
    eprint = "2302.03050",
    archivePrefix = "arXiv",
    primaryClass = "gr-qc",
    doi = "10.1103/PhysRevD.108.104020",
    journal = "Phys. Rev. D",
    volume = "108",
    number = "10",
    pages = "104020",
    year = "2023"
}

@article{Bhagwat:2019dtm,
    author = "Bhagwat, Swetha and Forteza, Xisco Jimenez and Pani, Paolo and Ferrari, Valeria",
    title = "{Ringdown overtones, black hole spectroscopy, and no-hair theorem tests}",
    eprint = "1910.08708",
    archivePrefix = "arXiv",
    primaryClass = "gr-qc",
    doi = "10.1103/PhysRevD.101.044033",
    journal = "Phys. Rev. D",
    volume = "101",
    number = "4",
    pages = "044033",
    year = "2020"
}

@article{LIGOScientific:2025rid,
    author = "Abac, A. G. and others",
    collaboration = "LIGO Scientific, Virgo, KAGRA",
    title = "{GW250114: Testing Hawking{\textquoteright}s Area Law and the Kerr Nature of Black Holes}",
    eprint = "2509.08054",
    archivePrefix = "arXiv",
    primaryClass = "gr-qc",
    reportNumber = "LIGO-P2500421",
    doi = "10.1103/kw5g-d732",
    journal = "Phys. Rev. Lett.",
    volume = "135",
    number = "11",
    pages = "111403",
    year = "2025"
}

@article{LIGOScientific:2025wao,
    author = "Abac, A. G. and others",
    collaboration = "LIGO Scientific, Virgo, KAGRA",
    title = "{Black Hole Spectroscopy and Tests of General Relativity with GW250114}",
    eprint = "2509.08099",
    archivePrefix = "arXiv",
    primaryClass = "gr-qc",
    reportNumber = "LIGO P2500461",
    doi = "10.1103/6c61-fm1n",
    journal = "Phys. Rev. Lett.",
    volume = "136",
    number = "4",
    pages = "041403",
    year = "2026"
}

@article{MaassenvandenBrink:2000iwh,
    author = "Maassen van den Brink, Alec",
    title = "{Analytic treatment of black hole gravitational waves at the algebraically special frequency}",
    eprint = "gr-qc/0001032",
    archivePrefix = "arXiv",
    reportNumber = "ITFA-2000-02",
    doi = "10.1103/PhysRevD.62.064009",
    journal = "Phys. Rev. D",
    volume = "62",
    pages = "064009",
    year = "2000"
}

@article{Chandrasekhar:1984mgh,
    author = "Chandrasekhar, Subrahmanyan",
    title = "{On algebraically special perturbations of black holes}",
    doi = "10.1098/rspa.1984.0021",
    journal = "Proc. Roy. Soc. Lond. A",
    volume = "392",
    number = "1802",
    pages = "1--13",
    year = "1984"
}

@article{Martel:2005ir,
    author = "Martel, Karl and Poisson, Eric",
    title = "{Gravitational perturbations of the Schwarzschild spacetime: A Practical covariant and gauge-invariant formalism}",
    eprint = "gr-qc/0502028",
    archivePrefix = "arXiv",
    doi = "10.1103/PhysRevD.71.104003",
    journal = "Phys. Rev. D",
    volume = "71",
    pages = "104003",
    year = "2005"
}

@article{Zenginoglu:2010cq,
    author = "Zenginoglu, Anil",
    title = "{Hyperboloidal layers for hyperbolic equations on unbounded domains}",
    eprint = "1008.3809",
    archivePrefix = "arXiv",
    primaryClass = "math.NA",
    doi = "10.1016/j.jcp.2010.12.016",
    journal = "J. Comput. Phys.",
    volume = "230",
    pages = "2286--2302",
    year = "2011"
}

@article{Bernuzzi:2011aj,
    author = "Bernuzzi, Sebastiano and Nagar, Alessandro and Zenginoglu, Anil",
    title = "{Binary black hole coalescence in the large-mass-ratio limit: the hyperboloidal layer method and waveforms at null infinity}",
    eprint = "1107.5402",
    archivePrefix = "arXiv",
    primaryClass = "gr-qc",
    doi = "10.1103/PhysRevD.84.084026",
    journal = "Phys. Rev. D",
    volume = "84",
    pages = "084026",
    year = "2011"
}

@article{Ma:2022wpv,
    author = "Ma, Sizheng and Mitman, Keefe and Sun, Ling and Deppe, Nils and H{\'e}bert, Fran{\c{c}}ois and Kidder, Lawrence E. and Moxon, Jordan and Throwe, William and Vu, Nils L. and Chen, Yanbei",
    title = "{Quasinormal-mode filters: A new approach to analyze the gravitational-wave ringdown of binary black-hole mergers}",
    eprint = "2207.10870",
    archivePrefix = "arXiv",
    primaryClass = "gr-qc",
    doi = "10.1103/PhysRevD.106.084036",
    journal = "Phys. Rev. D",
    volume = "106",
    number = "8",
    pages = "084036",
    year = "2022"
}

@article{Oshita:2025qmn,
    author = "Oshita, Naritaka and Ma, Sizheng and Chen, Yanbei and Yang, Huan",
    title = "{Probing Direct Waves in Black Hole Ringdowns}",
    eprint = "2509.09165",
    archivePrefix = "arXiv",
    primaryClass = "gr-qc",
    reportNumber = "YITP-25-141, RIKEN-iTHEMS-Report-25",
    month = "9",
    year = "2025",
    journal=""
}

@article{Kankani:2026kst,
    author = "Kankani, Anuj and McWilliams, Sean T.",
    title = "{The Direct Wave is Not a Meaningful Test of Horizon Properties}",
    eprint = "2607.02380",
    archivePrefix = "arXiv",
    primaryClass = "gr-qc",
    month = "7",
    year = "2026",
    journal=""
}

@article{Weller:2026wjh,
    author = "Weller, Colin and Laeuger, Andrew and Li, Fulin and Lu, Neil and Nathan, Rowina S. and Ma, Sizheng and Chen, Yanbei and Sun, Ling",
    title = "{From driven oscillations to free ringdown: a particle plunging into Kerr}",
    eprint = "2609.09531",
    archivePrefix = "arXiv",
    primaryClass = "gr-qc",
    month = "9",
    year = "2026",
    journal=""
}

@article{Su:2026fvj,
    author = "Su, Junquan and Khera, Neev and Casals, Marc and Ma, Sizheng and Chowdhuri, Abhishek and Yang, Huan",
    title = "{Decomposition of the Schwarzschild Green{\textquoteright}s function}",
    eprint = "2601.22015",
    archivePrefix = "arXiv",
    primaryClass = "gr-qc",
    doi = "10.1103/3n8r-8kbb",
    journal = "Phys. Rev. D",
    volume = "113",
    number = "10",
    pages = "104013",
    year = "2026"
}

@article{Ma:2026qbq,
    author = "Ma, Sizheng",
    title = "{Prompt Response from Plunging Sources in Schwarzschild Spacetime}",
    eprint = "2604.08680",
    archivePrefix = "arXiv",
    primaryClass = "gr-qc",
    month = "4",
    year = "2026",
    journal=""
}

@article{Chavda:2024awq,
    author = "Chavda, Ameya and Lagos, Macarena and Hui, Lam",
    title = "{The impact of initial conditions on quasi-normal modes}",
    eprint = "2412.03435",
    archivePrefix = "arXiv",
    primaryClass = "gr-qc",
    doi = "10.1088/1475-7516/2025/07/084",
    journal = "JCAP",
    volume = "07",
    pages = "084",
    year = "2025"
}

@article{Ori:2000zn,
    author = "Ori, Amos and Thorne, Kip S.",
    title = "{The Transition from inspiral to plunge for a compact body in a circular equatorial orbit around a massive, spinning black hole}",
    eprint = "gr-qc/0003032",
    archivePrefix = "arXiv",
    doi = "10.1103/PhysRevD.62.124022",
    journal = "Phys. Rev. D",
    volume = "62",
    pages = "124022",
    year = "2000"
}

@article{Szpak:2004sf,
    author = "Szpak, Nikodem",
    title = "{Quasinormal mode expansion and the exact solution of the Cauchy problem for wave equations}",
    eprint = "gr-qc/0411050",
    archivePrefix = "arXiv",
    month = "11",
    year = "2004",
    journal=""
}

@article{Detweiler:1979xr,
    author = "Detweiler, Steven L. and Szedenits, E.",
    title = "{Black holes and gravitational waves. II. Trajectories plunging into a nonrotating hole}",
    doi = "10.1086/157182",
    journal = "Astrophys. J.",
    volume = "231",
    pages = "211--218",
    year = "1979"
}

@misc{repo,
    author = "Kuntz, Adrien",
    title = "{Constant quasinormal-mode amplitudes for the ISCO plunge in Schwarzschild}",
    howpublished = "\url{https://github.com/akuntz00/QNM-amplitudes-ISCO-Schwarzschild}",
    year = "2026"
}

@article{Leung:2003eq,
    author = "Leung, P. T. and Maassen van den Brink, Alec and Mak, K. W. and Young, K.",
    title = "{Unconventional gravitational excitation of a Schwarzschild black hole}",
    eprint = "gr-qc/0301018",
    archivePrefix = "arXiv",
    doi = "10.1088/0264-9381/20/16/101",
    journal = "Class. Quant. Grav.",
    volume = "20",
    pages = "L217",
    year = "2003"
}

@article{Nagar:2005ea,
  archiveprefix = {arXiv},
  author = {Nagar, Alessandro and Rezzolla, Luciano},
  doi = {10.1088/0264-9381/23/12/C01, 10.1088/0264-9381/22/16/R01},
  eprint = {gr-qc/0502064},
  journal = {Class. Quant. Grav.},
  note = {[Erratum: Class. Quant. Grav.23,4297(2006)]},
  pages = {R167},
  primaryclass = {gr-qc},
  slaccitation = {%%CITATION = GR-QC/0502064;%%},
  title = {{Gauge-invariant non-spherical metric perturbations of Schwarzschild black-hole spacetimes}},
  volume = {22},
  year = {2005},
}

@misc{BHPToolkit,
  howpublished = {(\href{http://bhptoolkit.org/}{bhptoolkit.org})},
  title = {{Black Hole Perturbation Toolkit}},
}

@article{Zhang:2013ksa,
  archiveprefix = {arXiv},
  author = {Zhang, Zhongyang and Berti, Emanuele and Cardoso, Vitor},
  doi = {10.1103/PhysRevD.88.044018},
  eprint = {1305.4306},
  journal = {Phys. Rev. D},
  pages = {044018},
  primaryclass = {gr-qc},
  title = {{Quasinormal ringing of Kerr black holes. II. Excitation by particles falling radially with arbitrary energy}},
  volume = {88},
  year = {2013},
}

@article{LIGOScientific:2020tif,
  archiveprefix = {arXiv},
  author = {Abbott, R. and others},
  collaboration = {LIGO Scientific, Virgo},
  doi = {10.1103/PhysRevD.103.122002},
  eprint = {2010.14529},
  journal = {Phys. Rev. D},
  number = {12},
  pages = {122002},
  primaryclass = {gr-qc},
  reportnumber = {LIGO-P2000091},
  title = {{Tests of general relativity with binary black holes from the second LIGO-Virgo gravitational-wave transient catalog}},
  volume = {103},
  year = {2021},
}

@article{LIGOScientific:2019fpa,
  archiveprefix = {arXiv},
  author = {Abbott, B. P. and others},
  collaboration = {LIGO Scientific, Virgo},
  eprint = {1903.04467},
  journal = {a},
  primaryclass = {gr-qc},
  reportnumber = {LIGO-P1800316},
  slaccitation = {%%CITATION = ARXIV:1903.04467;%%},
  title = {{Tests of General Relativity with the Binary Black Hole Signals from the LIGO-Virgo Catalog GWTC-1}},
  year = {2019},
}

@article{Cotesta:2022pci,
  archiveprefix = {arXiv},
  author = {Cotesta, Roberto and Carullo, Gregorio and Berti, Emanuele and Cardoso, Vitor},
  doi = {10.1103/PhysRevLett.129.111102},
  eprint = {2201.00822},
  journal = {Phys. Rev. Lett.},
  number = {11},
  pages = {111102},
  primaryclass = {gr-qc},
  title = {{Analysis of Ringdown Overtones in GW150914}},
  volume = {129},
  year = {2022},
}

@article{Oshita:2024wgt,
  archiveprefix = {arXiv},
  author = {Oshita, Naritaka and Cardoso, Vitor},
  eprint = {2407.02563},
  month = {7},
  primaryclass = {gr-qc},
  reportnumber = {YITP-24-81, RIKEN-iTHEMS-Report-24},
  title = {{Reconstruction of ringdown with excitation factors}},
  year = {2024},
  journal=""
}

@article{Cheung:2023vki,
  archiveprefix = {arXiv},
  author = {Cheung, Mark Ho-Yeuk and Berti, Emanuele and Baibhav, Vishal and Cotesta, Roberto},
  eprint = {2310.04489},
  month = {10},
  primaryclass = {gr-qc},
  title = {{Extracting linear and nonlinear quasinormal modes from black hole merger simulations}},
  year = {2023},
  journal=""
}
\usepackage{aas_macros}
\usepackage{amsmath}
\usepackage{amssymb}
\usepackage{amsfonts}
\usepackage{physics}
\usepackage[normalem]{ulem}
\usepackage[english]{babel}
\usepackage[utf8]{inputenc}

\begin{document}

\title{Supplemental Material for ``The start time of quasi-normal modes"}
\author{Adrien Kuntz}
\affiliation{CENTRA, Departamento de Física, Instituto Superior Técnico – IST,
Universidade de Lisboa – UL, Avenida Rovisco Pais 1, 1049-001 Lisboa, Portugal}
\maketitle

\else
\newcommand{\SMend}{}
\clearpage
\onecolumngrid
\begin{center}\large\textbf{Supplemental Material}\end{center}
\twocolumngrid
\fi

\providecommand{\Ain}{A_{\rm in}}
\providecommand{\Aout}{A_{\rm out}}
\providecommand{\upk}{u_{\rm peak}}
\setcounter{equation}{0}\setcounter{figure}{0}\setcounter{table}{0}\setcounter{section}{0}
\setcounter{secnumdepth}{4}       
\renewcommand{\theequation}{S\arabic{equation}}
\renewcommand{\thefigure}{S\arabic{figure}}
\renewcommand{\thetable}{S\arabic{table}}
\renewcommand{\thesection}{S\arabic{section}}
\renewcommand{\theHequation}{S\arabic{equation}}  
\renewcommand{\theHfigure}{S\arabic{figure}}
\renewcommand{\theHtable}{S\arabic{table}}
\renewcommand{\theHsection}{S\arabic{section}}

\section{Orbit and time-domain waveform}\label{sec:orbit}

Here we give explicit expressions for the point-particle orbits and the source term of Regge-Wheeler and Zerilli (RWZ) equations. Since we consider a non-rotating BH, we can focus on the $\theta = \pi/2$ plane where the orbit satisfies
\begin{align}
    \frac{\mathrm{d}x_p}{\mathrm{d}t} &= - \frac{1}{E} \sqrt{E^2 - f(r) \bigg( 1 + \frac{L^2}{r^2}\bigg)} \label{eq:dxpdt} \; , \\
    \frac{\mathrm{d}\phi_p}{\mathrm{d}t} &= \frac{f(r) L}{E r^2} \; , \label{eq:dphipdt}
\end{align}
where $(x_p=r_p+\log(r_p-1), \phi_p)$ are the tortoise coordinate and azimuthal angle of the point-particle, $E$ and $L$ are its energy and angular momentum per unit mass, and $f(r)=1-1/r$. The plunge from the ISCO has $E=2\sqrt2/3$ and $L=2\sqrt3M$. With $w=\sqrt{r/(3-r)}$, and $t_p=\phi_p=0$ at the light ring $r=3M$, we can integrate Eqs.~\eqref{eq:dxpdt}-\eqref{eq:dphipdt} for the ISCO plunge to find
\begin{align}
t_p(r)&=-2\sqrt2(12-r)w+22\sqrt2\arctan w\nonumber\\
&\quad-\ln\frac{\sqrt2w-1}{\sqrt2w+1}-T_0 \; , \label{eq:tp}\\
\phi_p(r)&=2\sqrt3(1-w) \; ,\label{eq:phip}
\end{align}
with $T_0=-21\sqrt2+\tfrac{11}2\sqrt2\pi+2\ln(1+\sqrt2)\simeq-7.0M$. For convenience, we give the values of $t$, $u$ and $v$ along the orbit in Table~\ref{tab:events}.

\begin{table}[b]\caption{Events along the plunge from the ISCO. Times in units of $M$ from the peak of $|h_{22}|$, at
$\upk=-2.432M$ in the coordinates of Eq.~\eqref{eq:tp}.}\label{tab:events}
\begin{ruledtabular}\begin{tabular}{lcccc}
event & $r/M$ & $(t-\upk)/M$ & $(u-\upk)/M$ & $(v-\upk)/M$\\\hline
light ring & 3 & 2.43 & 0.82 & 4.05\\
bounce  & 2.557 & 6.06 & 6.06 & 6.06\\
horizon & 2 & $\infty$ & $\infty$ & 7.69\\
singularity & 0 & -- & -- & 9.43\\
\end{tabular}\end{ruledtabular}
\end{table}

Normalizing the point-particle mass to $\mu = 1$, the source functions $F$ and $G$ of the RWZ equations~\eqref{eq:rwz} are given by~\cite{Martel:2005ir}
\begin{align}
&\text{even sector:}\quad G=P\,(r^2+L^2),\qquad F=P\,B,\nonumber\\
&P=-\frac{16\pi\bar Yf}{rE\lambda L_r},\nonumber\\
&B=2imL \sqrt{R}+\frac12-\frac{\lambda r}2+\frac{(m^2-\lambda-1)L^2}r\nonumber\\
&\qquad+\frac{3(m^2-1)L^2}{(\lambda-2)r^2}+\frac{6E^2r}{L_r}\nonumber\\
&\qquad+\frac1{L_r}\Big[2(\lambda-2)L^2-\frac{3(\lambda-3)L^2}r\nonumber\\
&\qquad-\frac{6L^2}{r^2}-(\lambda+1)r\Big],\nonumber\\
R &= E^2 - f\bigg(1 + \frac{L^2}{r^2} \bigg) \nonumber \\
&\text{odd sector:}\quad G=\frac{QLf}{E^2}\Big(1+\frac{L^2}{r^2}\Big),\nonumber\\
&Q=\frac{16\pi\partial_\theta\bar Y}{r\lambda(\lambda-2)},\nonumber\\
&F=-\frac{QLf}{E^2r^4}\Big[2E^2r^3-r^3+\frac32r^2-2L^2r+\frac52L^2\nonumber\\
&\qquad-imLr^2 \sqrt{R}\Big],\label{eq:FG}
\end{align}
where $\lambda=\ell(\ell+1)$, $L_r=(\lambda-2)r+3$, $\bar Y=Y_{\ell m}(\tfrac\pi2,0)e^{-im\phi_p}$, and $\partial_\theta\bar Y=\partial_\theta Y_{\ell m}(\tfrac\pi2,0)e^{-im\phi_p}$. Solving the RWZ equations, one gets the strain at infinity with $r h_{\ell m}=\sqrt{(\ell+2)!/(\ell-2)!}\,\Psi_\mathrm{even}$ when $\ell+m$ is even and $r h_{\ell m} = i\sqrt{(\ell+2)!/(\ell-2)!}\,\Psi_\mathrm{odd}$ when $\ell + m$ is odd. 

We solve Eq.~\eqref{eq:rwz} in the time domain with hyperboloidal layers~\cite{Zenginoglu:2010cq,Bernuzzi:2011aj}. We use a uniform grid in $\rho$, where $\rho$ is defined as $x=\rho/\Omega(\rho)$ with $\Omega=1$ for $\rho<\rho_i=50M$ and $\Omega=1-[(\rho-\rho_i)/(\rho_\infty-\rho_i)]^4$ for $\rho>\rho_i$ (so that $\rho=\rho_\infty=150M$ is future null infinity). We similarly define the mirror layer for $\rho<-60M$ ($\rho=-110M$ is the horizon). We use fourth-order centered differences, fourth-order Runge--Kutta time stepping (Courant factor 0.75) and a grid spacing of $0.0125M$. The source is switched on smoothly over $-600M<t<-400M$, and $\delta$ and $\delta'$ are represented by Gaussians of width twice the grid spacing, whose effect is a smooth smearing of relative size $2\times10^{-5}$ (checked by halving the resolution).

\section{QNM data and mode functions}\label{sec:qnm}

In this Section we give details on our numerical procedure to compute the QNM frequencies $\omega_n$, the mode functions $u_\mathrm{in}$ and the ingoing amplitude $\Ain(\omega)$.

\textit{Frequencies.} Leaver's series for the Regge--Wheeler solution ingoing at the horizon $u_\mathrm{in}^\mathrm{RW}$ is~\cite{Leaver:1985ax}
\begin{equation}
u_{\rm in}^\mathrm{RW}=e^{-i\omega}(r-1)^{-i\omega}r^{2i\omega}e^{i\omega(r-1)}\sum_{k\geq0}b_k (1-1/r)^k, \label{eq:leaver}
\end{equation}
with $b_0=1$ and $\alpha_kb_{k+1}+\beta_kb_k+\gamma_kb_{k-1}=0$, $\alpha_k=(k+1)(k+1-2i \omega)$, $\beta_k=-[2k^2+(-8i\omega+2)k-8\omega^2-4i\omega+ \ell(\ell+1)-3]$, $\gamma_k=(k-2i\omega)^2-4$. Requiring the solution to converge at the QNM frequencies gives a continued-fraction equation that we solve to find the QNM frequencies $\omega_n$~\cite{Leaver:1985ax}, using the Nollert improvement~\cite{Nollert:1993zz}. Our values reproduces published results~\cite{BHPToolkit}; we note that the overtone $n=8$ of the $\ell=2$ harmonic ($\omega_8=-4i$) is an algebraically special mode, and it is not a zero of the ingoing amplitude $A_\mathrm{in}$~\cite{MaassenvandenBrink:2000iwh}.

\textit{Mode functions.} We evaluate $u_n$ and $\partial_xu_n$ from the series~\eqref{eq:leaver} at every point of the
orbit. It converges for $|1-1/r|<1$, which contains the whole orbit. Zerilli
solutions $u^\mathrm{Z}$ follow from the Chandrasekhar transformation~\cite{Chandrasekhar:1984mgh}
\begin{equation}
u^\mathrm{Z}=\frac{\mathcal A\,u^\mathrm{RW} +6\,\partial_x u^\mathrm{RW}}{\kappa_\ell-6i\omega} \; ,\quad\mathcal A=\kappa_\ell+\frac{18f}{r(\mu_\ell r+3)} \; ,
\label{eq:chandra}
\end{equation}
with $\kappa_\ell=(\ell-1)\ell(\ell+1)(\ell+2)$ and $\mu_\ell=(\ell-1)(\ell+2)$, which preserves the normalization at the
horizon and $\Ain$.

\textit{Excitation coefficients.} We compute $\Ain(\omega)$ with the complex scaling of Ref.~\cite{Lo:2025njp}. On the ray $x=x_0+\rho e^{i\beta}$, $\beta=-\arg\omega$, $\rho\geq0$, starting at a real point $x_0$, $e^{\pm i\omega x}=e^{\pm i\omega x_0}e^{\pm i|\omega|\rho}$ are two waves of constant modulus, so that $\Ain$ and $\Aout$ can be read off with comparable accuracy. We integrate the Regge-Wheeler equation along the ray, $d^2u/d\rho^2=e^{2i\beta}(V-\omega^2)u$, together with $dr/d\rho=e^{i\beta}f(r)$, which follows $r(x)$ continuously, with an eighth-order Runge--Kutta method, from initial data at $x_0$ given by Leaver's series (this is the only difference in our algorithm with respect to the one in~\cite{Lo:2025njp}, which started the integration from the horizon).
At large $|r|$ we match $u$ and $du/d\rho$ to the asymptotic series $e^{\pm i\omega x}\sum_{j\leq25}b^\pm_jr^{-j}$, where the coefficients $b_j$ satisfy a recursion relation found using the Regge-Wheeler equation: $\pm 2 i (j+1) \omega b_{j+1}^\pm = (j(j+1) - \ell(\ell+1) ) b_j^\pm - (j^2-4) b_{j-1}^\pm$.  The ray must pass the singular point $x(r=0)=i \pi $ on its right, $x_0>-\pi \cot\beta$. We can also find $\Ain$ on the imaginary axis by choosing $x_0=0.1>0$; this selects the branch $\Ain(-is + 0)$ on the negative imaginary axis, where $\Ain$ has a branch cut. 
The derivative $\Ain'(\omega_n)$ follows from central differences. Our values for $\Ain'(\omega_n)$ reproduces the ones in~\cite{Kuntz:2026xep}. For large $n$, $|\Ain'(\omega_n)| \simeq 20.4M$ tends to a constant.

\section{Constant amplitudes}\label{sec:const}

In this Section we give details on the technique we use to compute constant QNM amplitudes $\tilde C_n$. 

\textit{Source expansion.} The coefficients $a_k(\omega_n)$ of Eq.~\eqref{eq:ak} are the Taylor coefficients of $F^u\lambda_n^u-G^u\partial_x\lambda_n^u$ in $-U=e^{-u/2}$ along the orbit. We obtain them by expanding each factor in powers of $e^{-u/2}$ with power-series arithmetic. The first step is to obtain an expansion of the radial coordinate along the orbit. Using Eq.~\eqref{eq:dxpdt}, we get
\begin{equation}
    u(r) = u_\mathrm{LR} - \int_{r_\mathrm{LR}}^r \mathrm{d}r' \; \frac{E + \sqrt{R(r')}}{f(r') \sqrt{R(r')}} \; ,
\end{equation}
where $R(r)$ has been defined in Eq.~\eqref{eq:FG}. This can be expanded in a power series (the first coefficient being $u \simeq - 2 \log(r-1) + \mathcal{O}(1)$) and inverted to get $r-1$ as a function of $e^{-u/2}$ at any desired order. We then use Eq.~\eqref{eq:FG} to expand the source functions in $r-1$, Eq.~\eqref{eq:leaver} to expand $u_\mathrm{in}$ and $\partial_x u_\mathrm{in}$ in powers of $r-1$, and Eq.~\eqref{eq:dphipdt} to get $\phi_p$ in powers of $r-1$. 

The convergence radius of the series in $a_k(\omega_n)$ is fixed by the closest singularity of the source in this expansion. Besides $r=0$, there could be other singularities in the complex plane at which the source is singular, for example at $1-\dot x_p =0$ (see Eq.~\eqref{eq:sourceu}). However, we have checked by numerically computing the coefficients $a_k(\omega_n)$ for several orbits that, as expected, the convergence radius is indeed set by $r=0$. More precisely, we have checked that $a_k(\omega_n)/a_{k-1}(\omega_n) \to-e^{(v_S-u_B)/2}$ when the origin of time is at the bounce, which means that the start time is indeed given by Eq.~\eqref{eq:startTime}.

\textit{Vanishing of redshift modes and algebraically special frequency.} Since it is essential in the regularization procedure used in this article, let us briefly recap here the proof of Ref.~\cite{Kuntz:2026xep} showing that redshift modes vanish. The total waveform for $u>u_B$, including also the branch cut, can be found by convolving the full Green function~\eqref{eq:green} with the source and is given by
\begin{align}
    \Psi &= i \int \frac{\mathrm{d}\omega}{2 \pi} \frac{e^{-i \omega u}}{2 \omega A_\mathrm{in}(\omega)} \int_{-\infty}^u \mathrm{d}u' \, e^{i \omega u'} \nonumber \\
    &\times \bigg(  F^u  \lambda^u - G^u \partial_x \lambda^u\bigg) \; , \label{eq:psiLaplaceTransform}
\end{align}
with $\lambda^u = e^{i \omega x} u_\mathrm{in}$. 
For $u>u_S$, one can use a similar expansion to~\eqref{eq:ak} to obtain
\begin{align} \label{eq:vanishingRedshift}
    \Psi = \Psi_{u_1}(u) + i \sum_{k \geq 1} \int \frac{\mathrm{d}\omega}{2 \pi} \frac{e^{-i \omega u} a_k(\omega)}{2 \omega A_\mathrm{in}(\omega)} \bigg[ \frac{e^{(i \omega -k/2)u'}}{i \omega - k/2}  \bigg]_{u_1}^u ,
\end{align}
where $\Psi_{u_1}$ is the integral in Eq.~\eqref{eq:psiLaplaceTransform} with the upper bound set to $u=u_1$. 
The upper bound of the integral in Eq.~\eqref{eq:vanishingRedshift} vanishes because one can use Jordan lemma to close the contour of the $\omega$ integral in the upper half-plane, where no poles or branch cuts subsist~\cite{Kuntz:2026xep}. The lower bound is closed in the lower half-plane, where it picks up the usual QNM contribution along with the branch cut~\cite{Leaver:1986gd}. It may seem that a supplementary pole appears at the frequency $i\omega=k/2$ in Eq.~\eqref{eq:vanishingRedshift}, but this pole is canceled by a pole of $A_\mathrm{in}$ itself at this frequency~\cite{Kuntz:2026xep}.

This last statement, however, has one exception: at the algebraically special frequency $k=8$, $\Ain$ is regular (it has no zero, nor a pole, and is continuous across the branch cut there). This means that this frequency will show up in the spectrum: even if $n=8$ is not a QNM frequency, it appears in the time-domain signal through a redshift mode. To see this explicitly, notice that the upper bound of Eq.~\eqref{eq:vanishingRedshift} still vanishes by causality, so deforming the contour of integration gives
\begin{align}
    0 &= \sum_{n} \frac{a_8(\omega_n) e^{-4u}}{2 \omega_n \Ain'(\omega_n) (i \omega_n-4)} + \mathrm{B.C.} \nonumber \\
    &+ \frac{a_8(-4i) e^{-4u}}{8 \Ain(-4i) } \; , \label{eq:residueTheoremk=8}
\end{align}
where B.C. denotes the branch cut contribution. For $k=8$, neither the branch cut nor the additional pole contribution are small: consequently, the first term in Eq.~\eqref{eq:residueTheoremk=8} does not vanish any more. This term is exactly the upper bound in Eq.~\eqref{eq:ak} that was vanishing for $k \neq 8$: we thus see that the waveform contains a mode at $\omega = -4i$ whose amplitude is given by
\begin{equation}
    \Psi_\mathrm{QNM} \supset \bar C_8 e^{-4 u} \; , \quad \bar C_8 =\sum_{n} \frac{a_8(\omega_n)}{2 \omega_n \Ain'(\omega_n) (i \omega_n-4)} .
\end{equation}
The modulus of $\bar C_8$ is comparable to those of the overtones 7 and 9, as for an ordinary overtone. We added this algebraically special mode to the constant-amplitude model used in Fig.~\ref{fig:conv}.

\section{Branch cut and algebraically special mode}\label{sec:cut}

\begin{figure}[t]
\centering\includegraphics[width=\columnwidth]{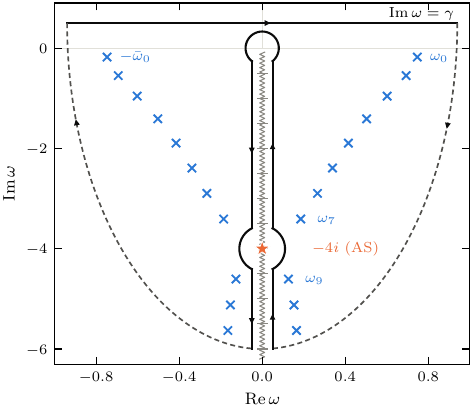}
\caption{Contour of integration of the Green function~\eqref{eq:green} in the complex frequency plane. For
$u\geq t'+|x'|$, the integral along ${\rm Im}\,\omega=\gamma$ is closed in the lower half-plane (dashed: arc at infinity).
It picks up the QNMs $\omega_n$ and their mirrors $-\bar\omega_n$ (crosses, $\ell=2$, $n\leq11$), and wraps around the
branch cut along the negative imaginary axis, with a detour around the pole of the Zerilli solution
$u_\mathrm{in}^\mathrm{Z}$ (even parity only) at the algebraically special frequency $\omega=-4i$ (star), where the
overtone $n=8$ is missing. Ticks: $\omega=-ik/2$, where $u_\mathrm{in}$ and $\Ain$ have poles that cancel each other in the Green function, except at $k=8$, where $\Ain$ is regular.}
\label{fig:contour}
\end{figure}

Here we numerically show that the residual of the waveform after subtracting the dynamical amplitude model, as seen in Fig.~\ref{fig:conv}(d), is completely given by the branch-cut integral and a dynamical QNM mode at an algebraically special frequency.

The Regge-Wheeler equation is regular at the algebraically special frequency $\omega\to-4i$, and its ingoing solution $u_\mathrm{in}^\mathrm{RW}$ can be found with Leaver series~\eqref{eq:leaver}. On the other hand, this is not the case for Zerilli equation: the Chandrasekhar transformation~\eqref{eq:chandra} shows that $u_\mathrm{in}^\mathrm{Z}$ develops a pole at $\omega=-4i$ (see also Ref.~\cite{MaassenvandenBrink:2000iwh}), with residue $(\mathcal{A} u_\mathrm{in}^\mathrm{RW} + 6 \partial_x u_\mathrm{in}^\mathrm{RW})/(-6i)$. This pole should be added when closing the contour of the Green function~\eqref{eq:green} in the lower half-plane for $u \geq t' + |x'|$ (Fig.~\ref{fig:contour}). This gives the following contribution to the Green function for the Zerilli equation:
\begin{equation}
    \mathcal{G}_\mathrm{AS} = - e^{-4(u-t')} \frac{\mathcal{A} u_\mathrm{in}^\mathrm{RW} + 6 \partial_x u_\mathrm{in}^\mathrm{RW}}{48 \Ain(-4i)} \; .
\end{equation}
The last remaining piece of the Green function when closing the integral in the lower half-plane is the branch cut along the negative imaginary axis, which from Eq.~\eqref{eq:green} can be written as
\begin{align}
    \mathcal{G}_\mathrm{cut} &= \int_0^\infty \mathrm{d}s\,c(s)\,e^{-s(u-t')}u_{\rm in}(x',-is) \; ,\nonumber\\
c(s)&=\frac{1}{4\pi(-is)}\Big[\frac1{\Ain(-is+0)}-\frac1{\Ain(-is-0)}\Big] \; ,\label{eq:cut}
\end{align}
where $\Ain(-is \pm 0)$ denotes the values of $\Ain$ on each side of the cut, and we have $\Ain(-is-0)=\overline{\Ain(-is+0)}$~\cite{Leung:2003eq}. Notice that, since $\Ain$ is real at the algebraically special frequency $s=4$, the pole in the Zerilli wavefunction $u_\mathrm{in}^\mathrm{Z}$ is canceled by a corresponding zero in $c(s)$ and the integral is convergent there. By convolving them with the source, we numerically evaluate the contribution of the cut and of the algebraically special mode to the waveform using the same techniques than in the rest of the paper. Since the tail depends on the whole past history of the system, in the branch cut integral we switch on the source functions $F$, $G$ at times $-600M<t<-400M$ with the same technique used in Sec.~\ref{sec:orbit}.

\begin{figure*}
\centering\includegraphics[width=\textwidth]{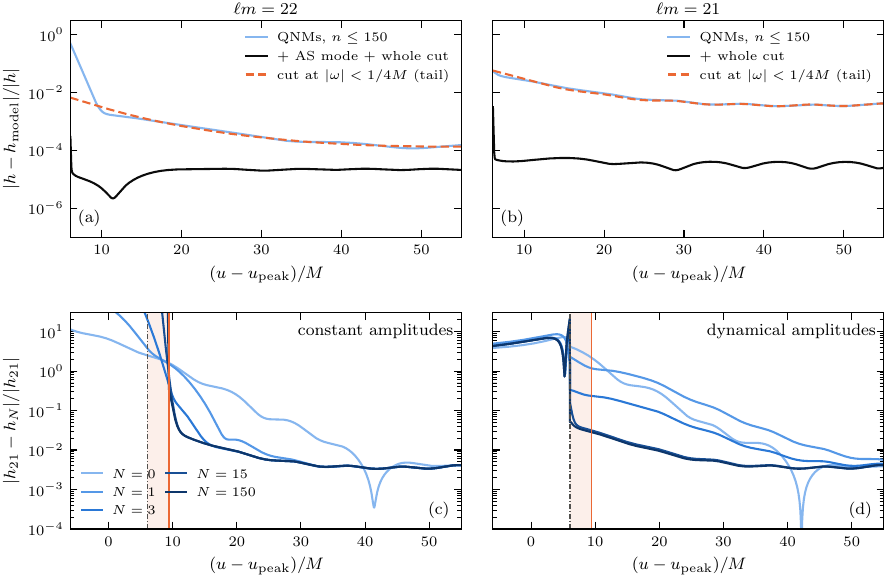}
\caption{(a),(b) Relative residual of the dynamical QNM sum after the bounce, alone and with the Zerilli mode at the
algebraically special frequency and the branch cut, for $\ell m=22$ and 21 (bounce at $\upk+6.06M$). Dashed: the
low-frequency part of the cut, $|\omega|<1/2$, i.e., the power-law tail, alone. (c),(d) As Fig.~\ref{fig:conv}(c),(d) for the
$\ell m=21$ mode.}\label{fig:beyond}
\end{figure*}

Fig.~\ref{fig:beyond}(a),(b) show what the dynamical QNM sum misses after the bounce. The residual of the dynamical QNM model with the Zerilli mode and the cut added is at the level of the error of the time-domain waveform, which validates our numerical procedure. 
The algebraically special mode is responsible for the somewhat large residual between the dynamical QNM model alone and the waveform at times $u_B \leq u \leq u_S$, which is apparent in Fig.~\ref{fig:conv}(d). On the other hand, since the 21 mode is odd, its Green function does not contain $\mathcal{G}_\mathrm{AS}$ and the residual should stay small as soon as $u \geq u_B$. We confirmed this behavior this by plotting the convergence of the two QNM sums (constant and dynamical) for $\ell m=21$ in Fig.~\ref{fig:beyond}(c),(d).

\SMend

\end{document}